\documentclass[a4paper,onecolumn,11pt,unpublished]{quantumarticle}
\pdfoutput=1
\usepackage[utf8]{inputenc}
\usepackage[english]{babel}
\usepackage[T1]{fontenc}
\usepackage{amsmath,amsfonts}
\usepackage{hyperref}
\usepackage{enumerate}
\usepackage{mhchem}
\usepackage[braket, qm]{qcircuit}
\usepackage{csvsimple}
\usepackage{booktabs}
\usepackage{supertabular}
\usepackage{siunitx}

\usepackage{tikz}
\usepackage{lipsum}

\begin{document}

\title{Potential energy surfaces inference of both ground and excited state 
using hybrid quantum-classical neural network}

\author{Yasutaka Nishida}
\email{yasutaka.nishida@toshiba.co.jp}
\affiliation{Corporate Research \& Development Center, 
Toshiba Corporation, 1 Komukai-Toshiba-cho, Saiwai-ku,
 Kawasaki 212-8582, Japan}
\author{Fumihiko Aiga}

\maketitle

\begin{abstract}
Reflecting the increasing interest in quantum computing, the 
variational quantum eigensolver (VQE) has attracted much 
attentions as a possible application of near-term quantum computers. 
Although the VQE has often been applied to quantum chemistry, 
high computational cost is required for reliable results because infinitely 
many measurements are needed to obtain an accurate expectation value and 
the expectation value is calculated many times to minimize a cost function 
in the variational optimization procedure. 
Therefore, it is necessary to reduce the computational cost of the 
VQE for a practical task such as estimating the potential energy surfaces 
(PESs) with chemical accuracy, which is of particular importance for 
the analysis of molecular structures and chemical reaction dynamics.
A hybrid quantum-classical neural network has recently been proposed 
for surrogate modeling of the VQE [Xia $et\ al$, \href{https://www.mdpi.com/1099-4300/22/8/828}
{Entropy \textbf{22}, 828 (2020)}]. Using the model, the ground 
state energies of a simple molecule such as \ce {H2} 
can be inferred accurately without the variational optimization procedure. 
In this study, we have extended the model by using the subspace-search variational 
quantum eigensolver procedure so that the PESs of the both ground and excited 
state can be inferred with chemical accuracy. We also demonstrate the effects 
of sampling noise on performance of the pre-trained model by using IBM's QASM 
backend.
\end{abstract}
%
%
\section{Introduction}
Together with recent advances in quantum hardware, the application field
of quantum computation has expanded to encompass physics 
\cite{phys1, phys2, phys3}, chemistry \cite{chem1,chem2,chem3}, 
machine learning \cite{qsv,math,math2,qml, qml2,qml3}, biology \cite{bio}, 
and finance \cite{fin}. Although promising applications of quantum computation 
for industry are still at the theoretical stage, the possibility of quantum advantage 
or required resource estimation for such applications is being actively studied 
\cite{qa1,qa2,qa3,qa4}. Quantum computers currently under development are called 
noisy intermediate-scale quantum (NISQ) \cite{nisq} devices. 
An NISQ has high noise level and can only perform imperfect operations with 
a limited coherence time. Therefore, many hybrid quantum-classical 
algorithms \cite{vqe,vqe2,qaoa, vqa1} have been developed to find 
an efficient use of this limited quantum resources.

The variational quantum eigensolver (VQE) \cite{vqe,vqe2} is one of the 
hybrid quantum-classical algorithms for solving electronic structure problems on 
quantum hardware. The VQE is a quantum heuristic algorithm, where a target expectation 
value is calculated based on the outcome from measurement on quantum hardware and 
parameters of a quantum circuit are updated to minimize the expectation value on a classical computer.
This variational optimization procedure is performed by iterating a closed feedback 
loop between classical and quantum hardware. To obtain accurate expectation values 
in the VQE procedure, a number of measurements and the averaging of their results are
required \cite{meas}. Therefore, for practical use, reduction techniques \cite{forging} 
for qubit and measurement \cite{sam1, quna2} 
have been actively studied. Furthermore, to extract a correct average value 
from noisy output or shorten computational runtimes, error 
mitigation techniques \cite{em1} and efficient optimization methods 
\cite{opt1,opt2} have been studied.

As regards other approaches for the practical use of quantum algorithms on NISQ devices, 
quantum machine learning has also been introduced \cite{qml3, quna, hqcnn, mvqe}. 
One such approach is applied to surrogate modeling or 
to building a generalized model to output VQE results in 
the field of quantum chemistry. For example, a Hamiltonian-alternating 
ansatz using a few training data points \cite{quna} or a hybrid quantum-classical 
neural network (HQCNN) model \cite{hqcnn, mvqe} are proposed. 
Since these generalized quantum circuits make it possible to infer the 
ground state energies accurately at any chemical configuration without 
the variational optimization procedure, calculation runtimes of PESs 
can be reduced drastically. 

In this paper, we have developed an HQCNN to estimate PESs of both ground 
and excited state, where the subspace-search variational quantum eigensolver 
(SSVQE) \cite{ssvqe} is used for minimizing a cost function. Our model 
is an extended model of the HQCNN proposed by Xia \cite{hqcnn}. A schematic 
approach is illustrated in Fig. \ref{fig1}. The difference between our work and 
Xia's is that we prepare mutually orthogonal initial states for calculating 
a cost function, and then the cost function is defined as a weighted sum 
of their energies. We apply our model to the \ce{H2} molecule and find that 
the pre-trained model can estimate PESs of both the ground and excited 
state with high accuracy. We also investigate the effects of sampling 
noise on performance of the pre-trained model by using IBM's QASM 
backend \cite{qiskit}, where our model is trained by a noiseless 
simulator and then PESs are inferred using the pre-trained model on 
IBM's qasm simulator. As a result, the relative errors between results from 
our model and the exact solution satisfy the chemical accuracy at $10^5$ QASM shots.

The paper is organized as follows. In Sec. \ref{sec2}, we briefly 
explain the procedure of VQE, SSVQE, and our HQCNN model. This is 
followed by the details of our implementation. In Sec. \ref{sec3}, 
we present our results and analyses that follow from our data. 
In Sec. \ref{sec4}, we discuss our results and outline open 
perspectives, and summarize in Sec. \ref{sec5}.
%
%
\section{Methods}
\label{sec2}
A surrogate model we have developed is explained in detail.  
Before proceeding to our model, we briefly review the VQE and the SSVQE.
\subsection{Variational quantum eigensolver}
\label{vqe}
The variational quantum eigensolver (VQE) \cite{vqe, vqe2} is a variational
algorithm that relies upon the Rayleigh-Ritz variational principle 
of quantum mechanics. A variational optimization procedure of 
the VQE minimizes the expectation value of a Hamiltonian by 
iterating a closed feedback loop between classical and quantum 
hardware. When a molecular Hamiltonian is given, the VQE can find 
its ground state energy and the ground state is generated on 
quantum hardware by an optimized parameterized quantum circuit, 
where the Hamiltonian is transformed into qubit representation 
from its second quantization representation. 
For an $n$-qubits system, a molecular Hamiltonian can be written 
as the sum of tensor products of Pauli matrices,
\begin{align}
\mathcal{H}=\sum_{P\in\{I,X,Y,Z\}^{\otimes n}}h_{P}P,
\label{eq1}
\end{align}
where $I$, $X$, $Y$, $Z$ are single-qubit Pauli operators and 
$h_p\in\mathbb{R}$ is a coefficient. The transformation of 
Eq. (\ref{eq1}) is done by the Jordan–Wigner transformation 
\cite{jwt1,jwt2}, the Bravyi–Kitaev transformation \cite{bkt1,bkt2} 
or other methods \cite{jwt1,bkt1,others}. The expectation 
value of the qubit Hamiltonian is evaluated on a classical 
computer by summing each expectation value of $P$ for 
$h_p\neq 0$ term which is the outcome from measurement on 
quantum hardware. To generate an initial quantum state in the 
VQE procedure, an initialized state $|0\rangle :=|0\rangle^{\otimes n}$ 
is fed into a parameterized quantum circuit $U(\vec{\theta})$.  
The classical parameters $\vec{\theta}$ of $U$ are updated 
to minimize the energy expectation value of 
$E(\vec{\theta})=\langle 0|U^{\dag}(\vec{\theta})\mathcal{H}U(\vec{\theta})|0\rangle$, 
where the updated parameters are fed back into the parameterized 
quantum circuit until $E(\vec{\theta})$ converges. 
When the minimal energy is reached, the optimal parameter 
$\vec{\theta}^*$ is determined. The VQE algorithm can be 
summarized as follows:
\begin{enumerate}[(i)]
\item Define a quantum circuit $U(\vec{\theta})$ with parameters $\vec{\theta}$.
\item Repeat the following procedures until $E(\vec{\theta})$ converges.
\begin{enumerate}[(a)]
\item Generate a quantum state $|\psi(\vec{\theta})\rangle=U(\vec{\theta})|0\rangle$.
\item Evaluate the energy $E(\vec{\theta})$ by measuring $\langle 0| U^{\dag}(\vec{\theta})
 \mathcal{H}U(\vec{\theta})|0\rangle$.
\item Update the parameters $\vec{\theta}$ to minimize $E(\vec{\theta})$.
\end{enumerate}
\end{enumerate}
When the convergence is reached, $E(\vec{\theta})$ is expected to 
be an approximate ground state energy.
\subsection{Subspace-search variational quantum eigensolver}
\label{ssvqe}
The subspace-search variational quantum eigensolver (SSVQE) \cite{ssvqe} 
is an extended algorithm of the VQE. The SSVQE utilizes the 
conservation of orthogonality under the unitary transformation. 
To search low-energy subspace of a given Hamiltonian, mutually orthogonal 
initial states are prepared and fed into a parameterized quantum circuit. 
The variational optimization is performed by minimizing a cost function defined 
as a weighted sum of energies of the mutually orthogonal quantum states.
The procedure of the SSVQE algorithm can be summarized as follows.

\begin{enumerate}[(i)]
\item Define a quantum circuit $U(\vec{\theta})$ and mutually orthogonal 
initial states (reference states) $\{|\phi_j\rangle \}_{j=0}^{k-1}$ where 
$\langle \phi_i|\phi_j\rangle=\delta_{ij}$ is satisfied. 
For a 4-qubit case, the computational basis $\{ |0000\rangle,\ |0001\rangle \}$ 
can be chosen as the two orthogonal reference states.
\item Repeat the following steps until a cost function is minimized.
\begin{enumerate}[(a)]
\item Generate mutually orthogonal quantum states $|\psi_j(\vec{\theta})\rangle
=U(\vec{\theta})|\phi_j\rangle$.
\item Evaluate a cost function defined as a weighted sum of energies of mutually orthogonal 
quantum states $|\psi_j(\vec{\theta})\rangle$, $L_{\omega}(\vec{\theta})=\sum_{j=0}^{k-1}\omega_j\langle \psi_j(\vec{\theta})|\mathcal{H}|\psi_j(\vec{\theta})\rangle$, 
where the vector of weight $\{\omega_j\}$ is chosen 
such that $\omega_0 > \omega_1 > \cdots >\omega_{k-1} >0$.
\item Update the parameters $\vec{\theta}$ to minimize $L_{\omega}$.
\end{enumerate}
\end{enumerate}
The vector of weight $\{\omega_j\}$ has the effect of adjusting which $|\phi_j\rangle$ converges 
to which excited state. When the cost function $L_{\omega}(\vec{\theta})$ converges,  
$U(\vec{\theta})$ maps the reference state $|\phi_j\rangle$ to the $j$th excited state 
$|\psi_j(\vec{\theta})\rangle$ of a given Hamiltonian. 
\subsection{Surrogate model of SSVQE}
\label{smodel}
\begin{figure}[h]
  \centering
  \includegraphics[scale=0.48]{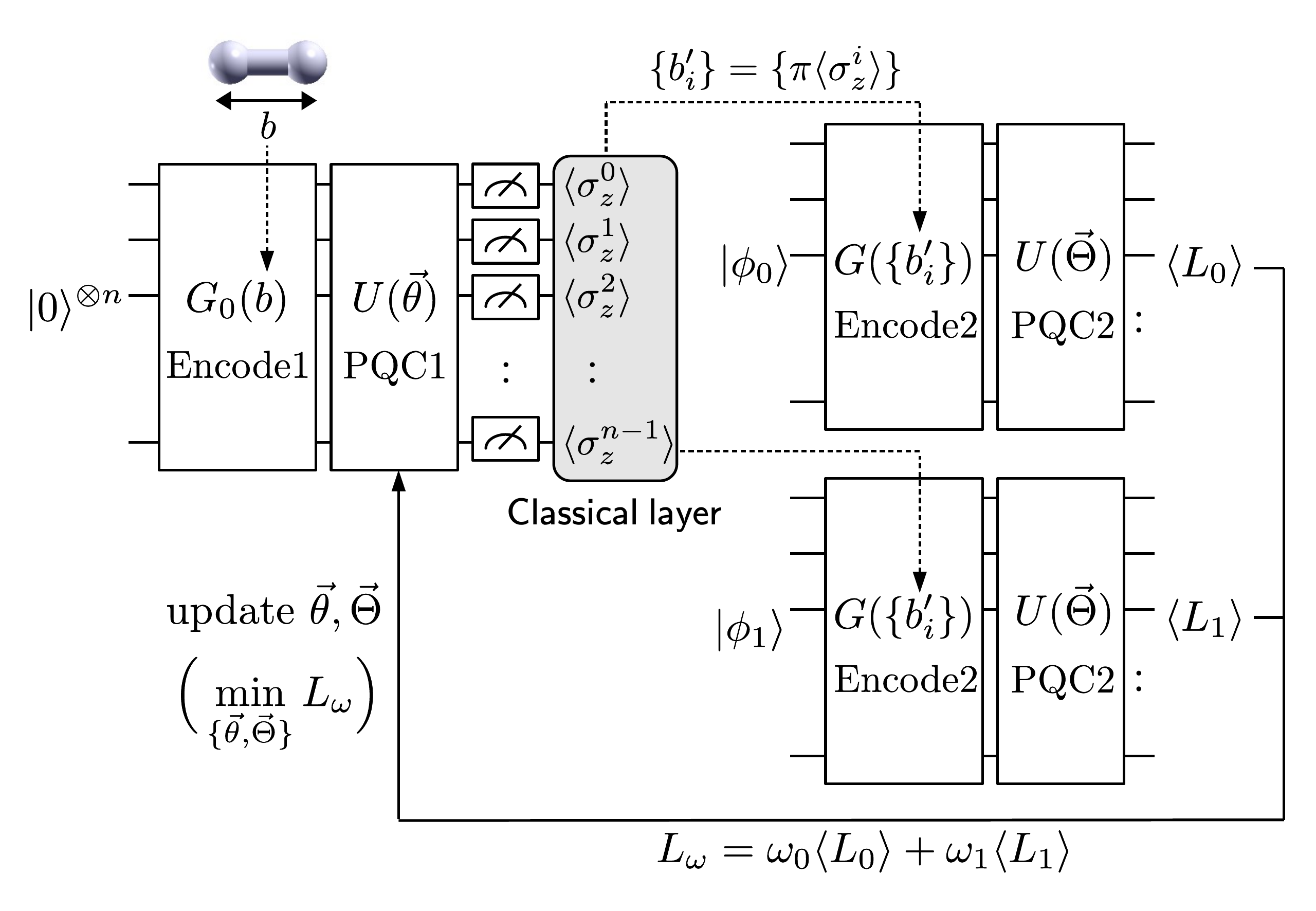}
  \caption{
  Our proposed extended quantum-classical hybrid neural network. 
  Encode1 and Encode2 represent encode layers, and PQC1 and PQC2 represent 
  parameterized quantum circuits $U$. 
  In the first layer, bond length $b$ is given as an input parameter 
  of Encode1, $G_0$, and then each qubit measurement is performed. 
  The expectation value of Pauli-Z matrices $\langle\sigma_z^i\rangle$ 
  calculated from $i_{th}$ qubit measurement are fed into Encode2 as 
  an input $\{b'_i\}=\{\pi \langle\sigma_z^i\rangle\}$ of $G$.  $|\phi_0\rangle$ 
  and $|\phi_1\rangle$ represent orthogonal reference states that 
  finally converges to ground state and 1st excited state, respectively.
  $\langle L_0\rangle$ and $\langle L_1\rangle$ are the summation of the 
  expectation value of each Hamiltonian calculated from training data 
  points for each reference state $|\phi_0\rangle$ and $|\phi_1\rangle$, 
  respectively. The parameters $\vec{\theta}$ and $\vec{\Theta}$ 
  in PQC1 and PQC2 are optimized to minimize cost function 
  $L_{\omega}=\omega_0 \langle L_0(\{\vec{\theta},\vec{\Theta} \}) \rangle
  +\omega_1 \langle L_1(\{ \vec{\theta},\vec{\Theta} \})\rangle$.
  Graphics of \ce{H2} molecule is generated by XCrySDen \cite{xcrys}.
  } 
  \label{fig1}
\end{figure}

A surrogate model aims to generalize the outputs from a particular 
operation and reduce its computational cost. In the previous 
study \cite{hqcnn}, a hybrid quantum-classical neural network 
(HQCNN) is proposed to generalize outputs from a circuit optimized 
by the VQE. The HQCNN model can be regarded as a combining network 
between linear operations of quantum circuits and nonlinear operations of 
measurements and can infer the PES of the ground state with high accuracy 
for simple molecules such as \ce{H2}, \ce{LiH}, and \ce{BeH2} \cite{hqcnn}. 
In this study, we extend the original HQCNN model 
so that both the ground and excited state energies can be inferred at once. 
Our strategy is as follows. In the original model, a VQE-based procedure is used as 
unsupervised training for optimizing the quantum-classical neural network, where the 
energy summation of each input data point is minimized. In our model, 
the VQE-based training procedure is replaced with an SSVQE one. 
Fig. \ref{fig1} shows our extended HQCNN model for two orthogonal 
reference states of $|\phi_0\rangle$ and $|\phi_1\rangle$. 

The details of implementation can be summarized as follows.
\subsubsection{Data encoding}
\label{sec3-1}
The first step is to encode the input classical data into a quantum state.
On the basis of the previous work \cite{hqcnn}, the encode gate $G$ for an 
$n$-qubits system is generally defined as, $G=\otimes_{i=0}^{n-1}g_i(f_i(a_i))$
where $g_i$ is a set of single qubit quantum gates on qubit $i$ and $f_i$ 
is a classical function to encode $a_i$ as the parameter of $g_i$. The encode 
circuit $G$ is initialized in the $|0\rangle^{\otimes n}$. 
An example of $G$ is written as follows,
\begin{align}
G(\{a_i\})=\otimes_{i=0}^{n-1}R_y(a_i)H,
\label{eq2}
\end{align}
where $R_y$ is the rotation-$y$ gate and $H$ is the 
Hadamard gate. In this work, we focus on a simple diatomic molecule, \ce{H2}. 
Therefore, the bond length $b$ is set as input data, 
so that the encode gate $G_0$ having one input parameter 
can be written as follows, 
\begin{align}
G_0(b)=\otimes_{i=0}^{n-1}R_y(b)H.
\label{eq3}
\end{align}
\subsubsection{Parameterized Quantum Circuit}
\label{sec3-2}
\begin{figure}[h]
  \centering
  \includegraphics[scale=0.52]{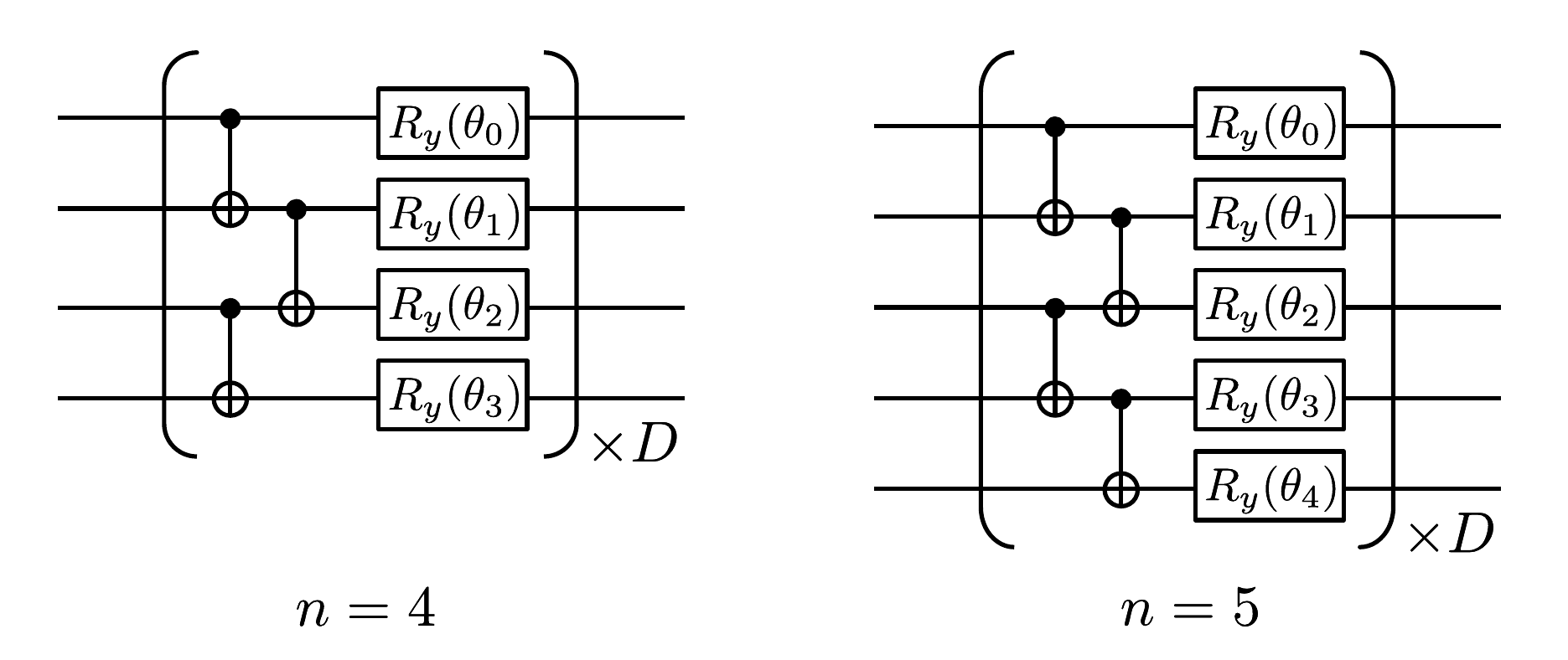}
  \caption{Examples of a PQC based on RealAmplitudes circuits with $D$ depth 
  in $n=4$, $5$ cases.
  } 
  \label{fig2}
\end{figure}

The second step is to build a parameterized quantum circuit (PQC) that 
consists of individual qubit rotations and qubit entangling parts. 
As all quantum gates used in a quantum circuit are unitary, a PQC 
itself is described as a unitary operation on $n$-qubits, $U(\vec{\theta})$.
Since the expressibility and entangling capability of a PQC depend on its 
circuit structure \cite{pqc}, design of a circuit structure is 
important for inference performance of the HQCNN. Here, we use 
a simple circuit of RealAmplitudes comprising alternating 
$R_y$ rotation gates on a single qubit and controlled-X ($CX$) gates. 
For the $n$-qubit case, the PQC with circuit depth $D(\ge 1)$ is expressed by 
\begin{align}
&\text{$\bullet$ $n$ is even}\nonumber\\
&\ \ \ \ \ U(\vec{\theta})=\prod_{d=1}^{D}(\otimes_{i=0}^{n-1}R_y(\theta_{i+n(d-1)}))
(CX_{n-3,n-2}\cdots CX_{1,2})(CX_{n-2,n-1}\cdots CX_{0,1}),\label{eq4}\\
&\text{$\bullet$ $n$ is odd}\nonumber\\
&\ \ \ \ \ U(\vec{\theta})=\prod_{d=1}^{D}(\otimes_{i=0}^{n-1}R_y(\theta_{i+n(d-1)}))
(CX_{n-2,n-1}\cdots CX_{1,2})(CX_{n-3,n-2}\cdots CX_{0,1}),
\label{eq5}
\end{align}
where $CX_{m,n}$ represents $CX$ gate with $m$ as the control qubit and 
$n$ is the target qubit. Examples of $U(\vec{\theta})$ in $n=4$ and 5 cases 
are illustrated in Fig. \ref{fig2}. Here, the depth $D$ is defined by the 
repetition number of the block circuit as shown in Fig. \ref{fig2}.
For $U(\vec{\theta})$ with $n$-qubits and $D$ depth, 
the number of the parameters is $nD$, 
$\vec{\theta}=\{\theta_0,\theta_1,\cdots \theta_{nD-1}\}$.
\subsubsection{Classical layer}
\label{sec3-3}
In Fig. \ref{fig1}, the measurement followed by PQC1 is a classical layer 
that corresponds to the activation function connecting between quantum layers. 
The classical layer measures the expectation value of Pauli-Z matrices of each 
qubit $i$, $\langle \sigma_z^i\rangle$. The values of 
$\langle \sigma_z^i\rangle$ are fed into the second encode
layer as $G(\{\pi\langle \sigma_z^i \rangle\})$, and then the reference states 
$|\phi_j\rangle$ are encoded into a quantum state.
\subsubsection{Cost function}
\label{sec3-4}
A cost function is defined as a weighted energy summation of each reference state,
\begin{align}
L_\omega(\vec{\theta},\vec{\Theta})=\sum_{j=0}^{k-1}\omega_j \langle L_j \rangle,
\label{eq6}
\end{align}
where $\langle L_j \rangle$ is the summation of the expectation value of each Hamiltonian 
calculated from training data points for each reference state $|\phi_j\rangle$. 
As explained in Sec. \ref{ssvqe}, the relation of 
$\omega_0 > \omega_1 >\cdots >\omega_{k-1}>0$ is satisfied. 
When the training points are chosen as ${\vec{b}}=\{b_0,b_1,\cdots b_{\alpha-1}\}$, 
$\langle L_{j}\rangle$ is calculated by
\begin{align}
&\langle L_j \rangle = \frac{1}{\alpha}\sum_{m=0}^{\alpha-1}
\langle \psi_{j}(b_m)|U^{\dag}(\vec{\Theta})\mathcal{H}(b_m)U(\vec{\Theta}) |\psi_{j}(b_m)\rangle,\label{eq7}\\
&|\psi_{j}(b_m)\rangle =G(\{\pi\langle \sigma_{z,m}^i\rangle\})|\phi_j\rangle,\label{eq8}\\
&\langle \sigma_{z,m}^i\rangle=\langle \psi(b_m) |U^{\dag}(\vec{\theta})
(I\otimes I \cdots I \otimes \underbrace{Z}_{i\rm{th}}
\otimes I \cdots I \otimes I)U(\vec{\theta}) |\psi(b_m)\rangle,\label{eq9}\\
&|\psi(b_m)\rangle =G_0(b_m)|0\rangle\label{eq10}.
\end{align}
For the $n$-qubits PQC with $D$ depth, the cost function of Eq. (\ref{eq6}) 
is a function of $2nD$-dimensional vectors , i.e., $L_{\omega}(\{ 
\theta_0,\theta_1,\cdots \theta_{nD-1}\},\{ 
\Theta_0,\Theta_1,\cdots \Theta_{nD-1}\})$.
\label{eq11}
In the $k=2$ case, the cost function is given by 
\begin{align}
L_\omega(\vec{\theta},\vec{\Theta})= \omega_0 \langle L_0 \rangle 
+ \omega_1 \langle L_1 \rangle.
\label{eq11}
\end{align}
When the weight $\omega_j$ is fixed as $(\omega_0,\omega_1)=(1,0)$, 
our model corresponds to Xia's original model \cite{hqcnn} which can estimate 
only ground state energies. For estimating both energies of ground state and 1st 
excited state, we only have to set an appropriate weight such 
as $(\omega_0,\omega_1)=(1,0.5)$ to satisfy $\omega_0 > \omega_1$. 
To check validity of our implementation, the trivial case of the 
$(\omega_0,\omega_1)=(1,0)$ model is discussed in Sec. \ref{sec3}.

\subsubsection{Optimization and computational conditions}
\label{sec3-5}
The parameters $\vec{\theta}$ and $\vec{\Theta}$ of the PQC are updated to 
minimize the cost function $L_\omega$. 
The optimization is performed by the Broyden–Fletcher–Goldfarb–Shanno 
algorithm \cite{bfgs} provided in the SciPy library \cite{scipy}.
The entire simulation of the quantum circuit is performed by Qiskit \cite{qiskit}, 
where the statevector or backend noise model (QASM) simulator is available. 
The computational conditions are set as maximum iterations and gradient norm 
tolerance of 1000 and $10^{-5}$, respectively. The qubit Hamiltonian of 
the equation (\ref{eq1}) at each chemical configuration is transformed 
from its second quantization representation by OpenFermion \cite{ofermi}, 
where the second quantization Hamiltonian is calculated by STO-3G 
minimal basis using PySCF \cite{pyscf}. The transformation of Eq. (\ref{eq1})
is done by the Jordan–Wigner transformation \cite{jwt1,jwt2}.
%
%
\section{Numerical simulation and results}
\label{sec3}
Using our HQCNN model, PES inference for the \ce{H2} molecule is performed. 
In this simulation, each Hamiltonian of \ce{H2} 
is transformed into 4-qubit Hamiltonian ($n=4$) while varying its bond length. 
To train the PQC, a few training data points are used as 
$b_{\rm{train}}=\{ 0.45, 0.85, 1.25, 1.65, 2.05, 2.45\}$ $\rm{\AA}$. 
The parameters of $U$ are optimized on the statevector simulator 
in Qiskit \cite{qiskit}, i.e., a noiseless simulator. Other computational 
conditions are summarized in Sec. \ref{sec3-5}.

\subsection{PES inference for ground state}
\label{sec3-1}
At first, the $(\omega_0,\omega_1)=(1,0)$ case is performed 
to check the validity of our implementation. 
The depth $D$ of the PQC, $U(\vec{\theta})$ or $U(\vec{\Theta})$, 
is varied from 2 to 6. Hence, the total number of parameters can 
be $2nD=16\sim 48$. The optimal parameters $(\vec{\theta}^*,\vec{\Theta}^*)$ 
for each depth model are summarized in appendix \ref{appenA}. 
Once the optimal parameters $\vec{\Theta}^*$ are determined, 
the ground state energies at any bond length $b_{\rm{test}}=\{b_l \}$ 
are inferred as follows,
\begin{align}
E_0(b_l) = \langle \psi_{0}(b_l) | U^{\dag}(\vec{\Theta}^*)\mathcal{H}(b_l)U(\vec{\Theta}^*)|
\psi_{0}(b_l)\rangle,
\label{eq12}
\end{align}
where 
$|\psi_{0}(b_l)\rangle = G(\{\pi\langle \sigma_{z,l}^i\rangle)\}|\phi_0\rangle$ 
as shown in the equation (\ref{eq8}).

\begin{figure}[h]
  \centering
  \includegraphics[scale=0.57]{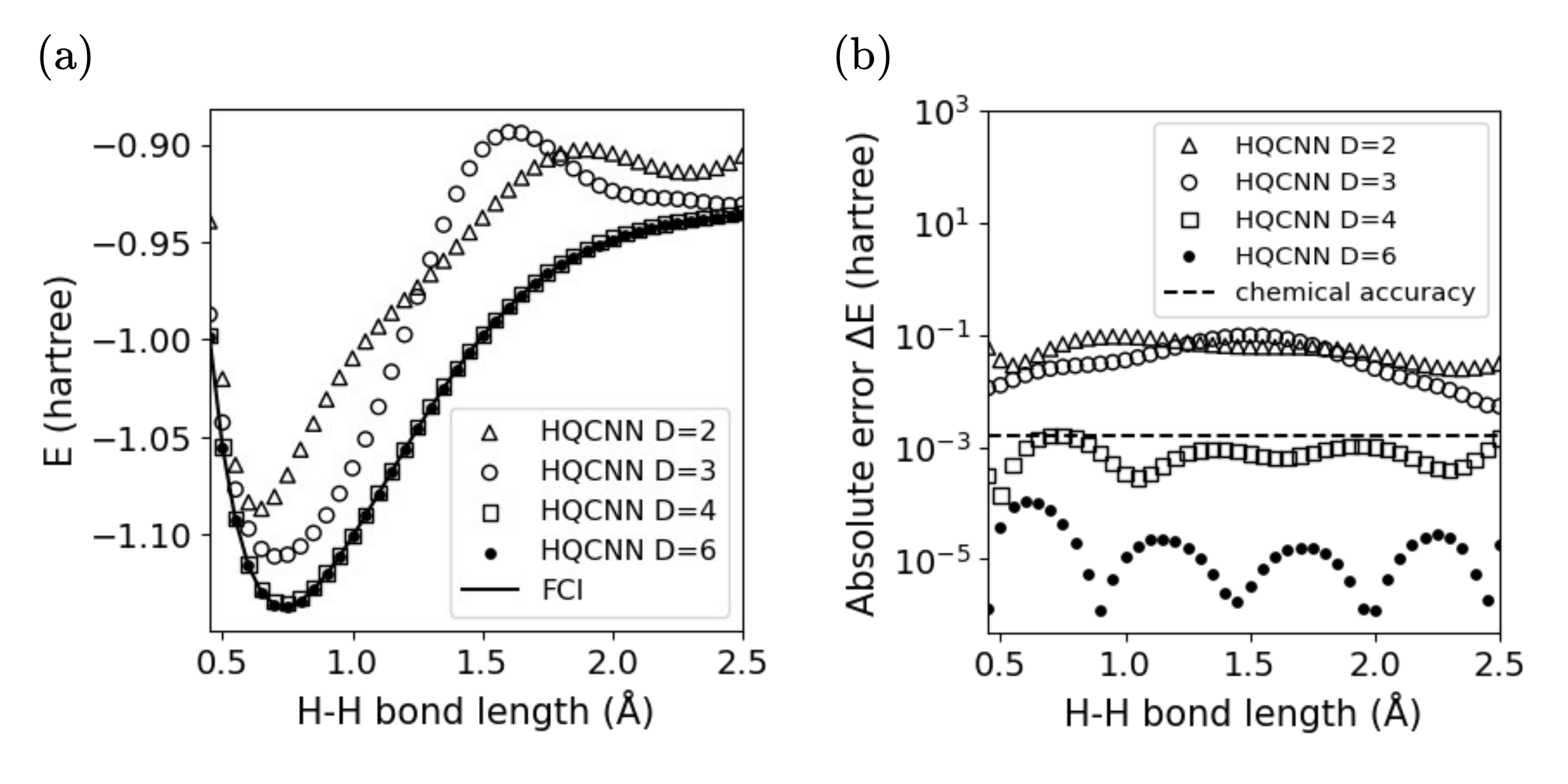}
  \caption{(a) PES inference for \ce{H2} molecule using the $(\omega_0,\omega_1)=(1,0)$ 
  HQCNN model with each depth $D$. The solid line in the figure is drawn 
  by full configuration interaction (FCI) calculation. (b) Absolute 
  errors of the energies between each HQCNN model and FCI. The dotted 
  line in the figure is drawn from the value of chemical accuracy, 
  0.001593 hartree.
  } 
  \label{fig3}
\end{figure}

Figure \ref{fig3}(a) shows the results of PES inference of the ground state 
and full configuration interaction (FCI) calculation.  
One can clearly see the results of $D\ge 4$ match well the exact 
solution of the FCI method. Figure \ref{fig3}(b) shows the relative errors 
between each HQCNN model and the FCI method. The error in the case of $D=6$ 
does not exceed the chemical accuracy, $\sim 1.6\cdot 10^{-3}$ hartree.
Hence, $D=6$ is considered to have enough expressibility to 
describe the \ce{H2} system.

To confirm consistency with Xia's work \cite{hqcnn}, we have also 
calculated the PES without intermediate measurements, where full linear operation 
is performed by removing the classical layer. 
These results are summarized in Appendix \ref{appenB}. From our calculation, 
the absence of a classical layer make inference accuracy worse. 
This is consistent with Xia's work and one can see that nonlinear connections 
between linear unitary operations are essential for highly 
accurate PES inference over the whole range of the bond length.

\subsection{PES inference for both ground and excited state}
\label{sec3-2}
The $D=6$ condition is expected to have enough expressibility for the \ce{H2} 
system described in Sec. \ref{sec3-1}. Next, the results of the 
$(\omega_0,\omega_1)= (1, 0.5)$ model with $D=6$ are shown here. 
In this case, the parameters of the PQC are optimized to minimize the cost 
function $L_{\omega}=\langle L_0\rangle +0.5\langle L_1\rangle$ 
from the equation (\ref{eq11}). By using the equation (\ref{eq7}) for optimal parameters 
$(\vec{\theta}^*,\vec{\Theta}^*)$, the ground state energies $E_0$ and the excited state 
energies $E_1$ at any bond length $b_{\rm{test}}=\{b_l\}$ are inferred by
\begin{align}
E_j(b_l) &= \langle \psi_{j}(b_l) | U^{\dag}(\vec{\Theta}^*)\mathcal{H}(b_l)U(\vec{\Theta}^*)|
\psi_{j}(b_l)\rangle \ (j=0,1),\label{eq13}
\end{align}
where 
$|\psi_{j}(b_l)\rangle = G(\{\pi\langle \sigma_{z,l}^i\rangle\})|\phi_j\rangle$ 
for two reference states $\phi_0=|0000\rangle$ and  $\phi_1=|0001\rangle$ 
(little-endian for qubit ordering is used here.). 
The optimal parameters $(\vec{\theta}^*,\vec{\Theta}^*)$ are summarized in Appendix \ref{appenC}. 

Figure \ref{fig4}(a) shows the PES results from $E_0$ and $E_1$ inferred by the HQCNN model. 
Compared with the FCI method, the results of both states match well. Figure \ref{fig4}(b) 
shows the relative errors between results using the HQCNN model 
and the FCI method. We find that the error satisfies the chemical accuracy 
in both states and the PES inference for the excited state is slightly more accurate.
\begin{figure}[h]
  \centering
  \includegraphics[scale=0.57]{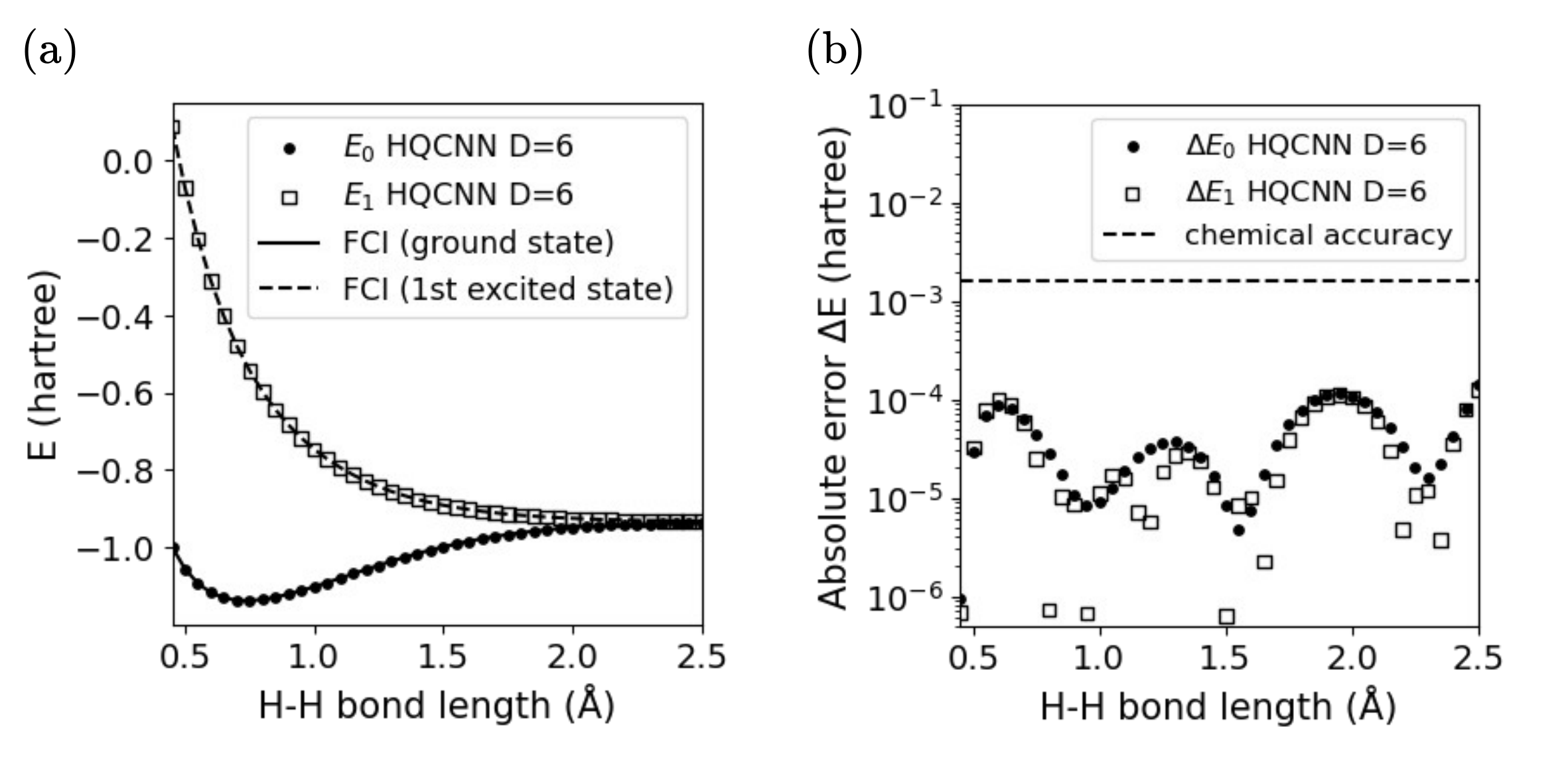}
  \caption{(a) PES inference of \ce{H2} molecule using the $(\omega_0,\omega_1)=(1,0.5)$ 
  HQCNN model with $D=6$. The solid and dashed lines in the figure are drawn by full 
  configuration interaction (FCI) calculation. 
  (b) Absolute errors between HQCNN model with $D=6$ and FCI for $E_0$ and $E_1$. 
    The dashed line in the figure is drawn from the value of chemical accuracy, 
    0.001593 hartree.
  } 
  \label{fig4}
\end{figure}

\subsection{Effect of sampling noise on PES inference}
\label{sec3-3}
The results discussed so far are obtained on the statevector simulator, 
i.e., without noise due to sampling. On actual quantum hardware, the 
statevector cannot be observed and the noise effect from measurements 
is unavoidable. Here, focusing on the sampling noise only, 
the noise effect on PES estimation is simply discussed as follows: 
the parameters $(\vec{\theta}^*,\vec{\Theta}^*)$ of the HQCNN model 
are predetermined on the statevector simulator, and then the energy 
estimation is performed by using the optimized HQCNN model 
on IBM's qasm simulator \cite{qiskit}. 

\begin{figure}[h]
  \centering
  \includegraphics[scale=0.57]{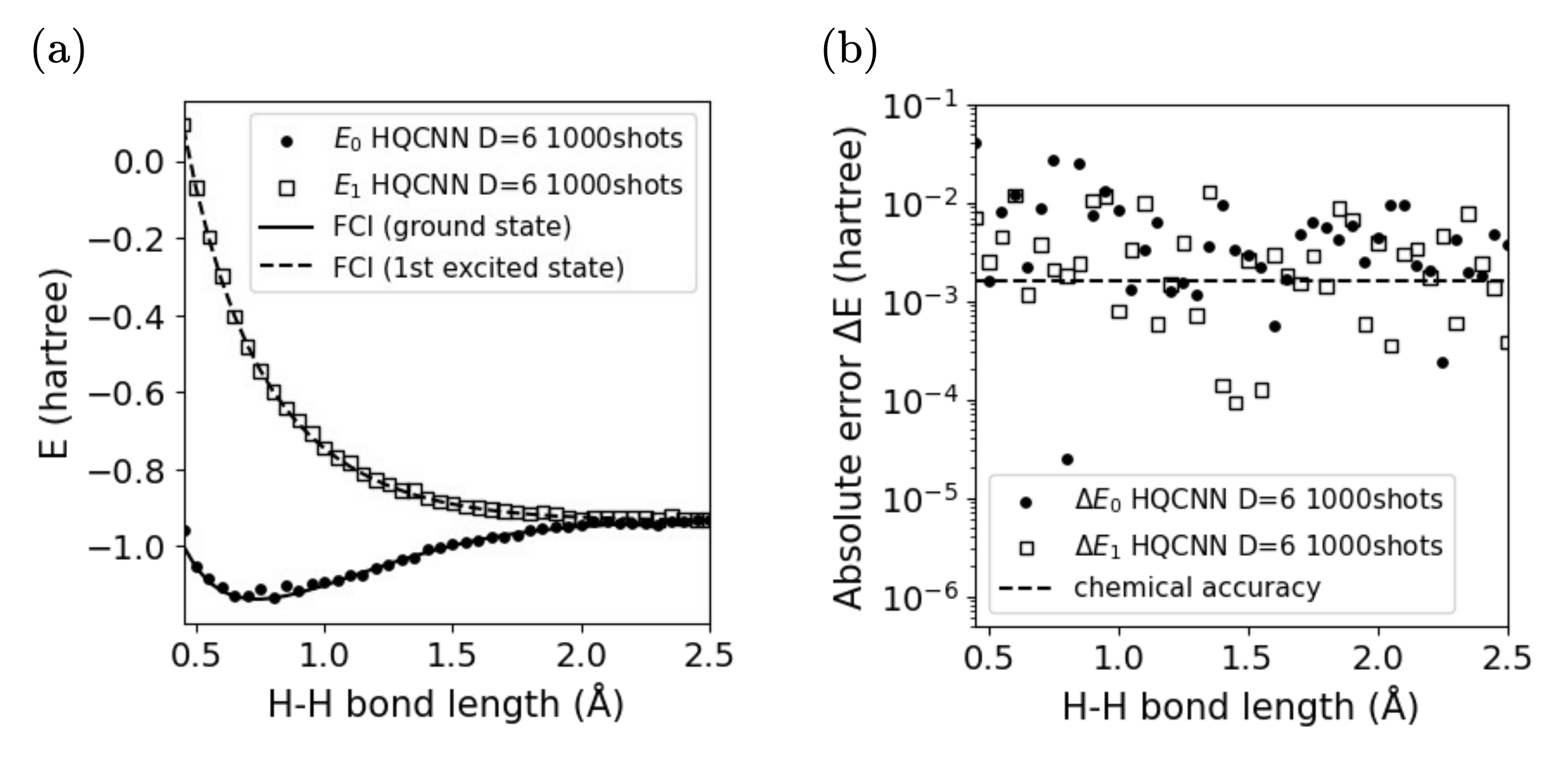}
  \includegraphics[scale=0.57]{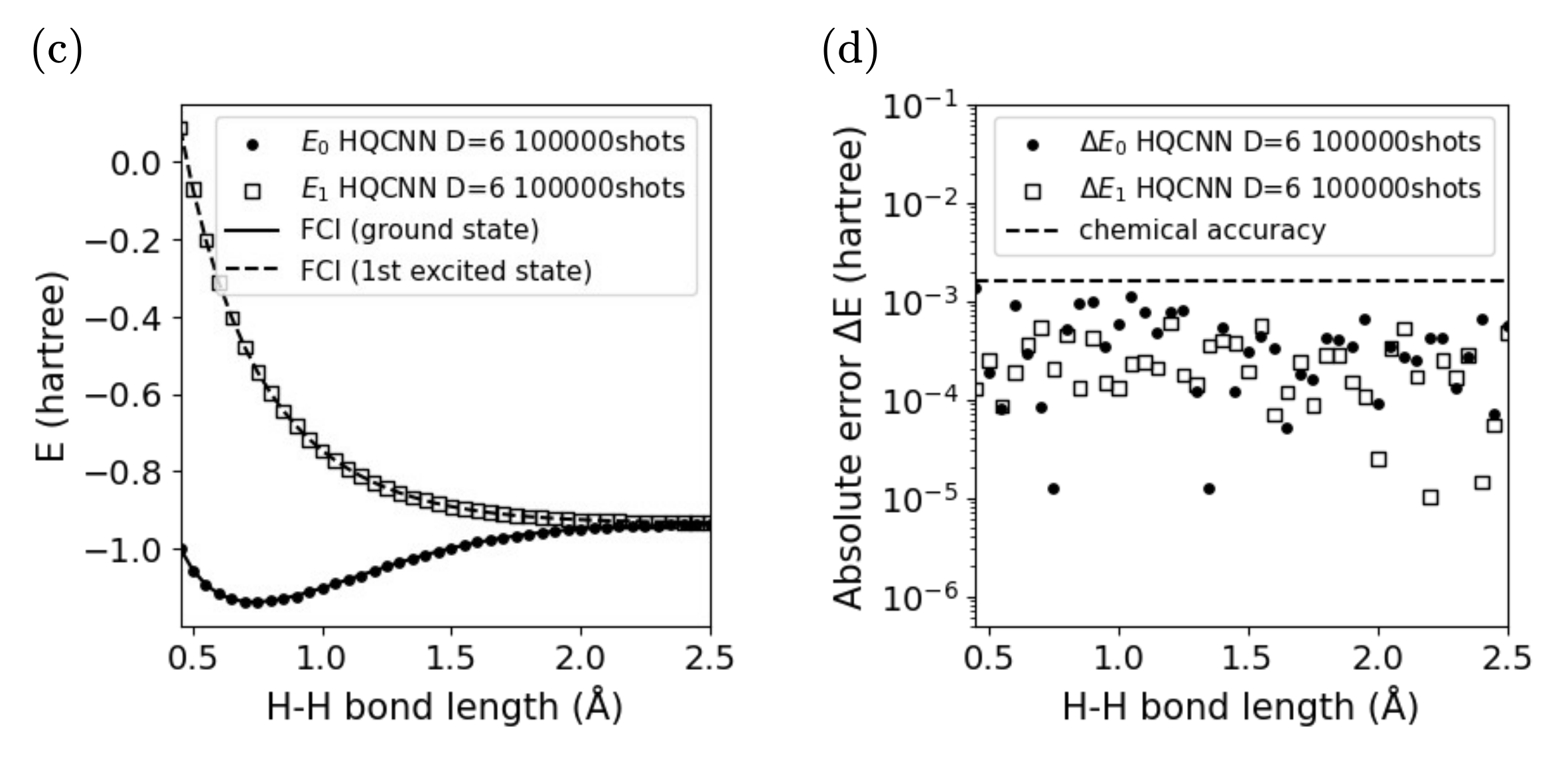}
  \caption{Sampling noise effects on PES inference of the 
  $(\omega_0,\omega_1)=(1,0.5)$ HQCNN model, where the expectation value of the energies 
  are estimated on the qasm simulator. (a) PES 
  of \ce{H2} estimated by an HQCNN model with 1000 shots. 
  (b) Absolute errors of the energies between the HQCNN model and FCI in the panel (a).
  (c) PES of \ce{H2} estimated by an HQCNN model with 100000 shots. 
  (d) Absolute errors of the energies between the HQCNN model and FCI in the panel (c). 
  The dashed lines in the panel (b) and (d) are drawn from 
  the value of chemical accuracy, 0.001593 hartree.
  } 
  \label{fig5}
\end{figure}

Figure \ref{fig5} shows the influence of shot number on PES inference. 
In the panel of Fig. \ref{fig5}(a) and (c), the PESs are plotted 
in the case of $10^3$ and $10^5$ shots. In the panel of Fig. \ref{fig5}(b) 
and (d), the energy differences $\Delta E_0$ and $\Delta E_1$ 
between the HQCNN model and the FCI method are displayed for each shot 
condition. One can see clearly that the errors eventually decrease as 
the shot number increases.

\begin{figure}[h]
  \centering
  \includegraphics[scale=0.57]{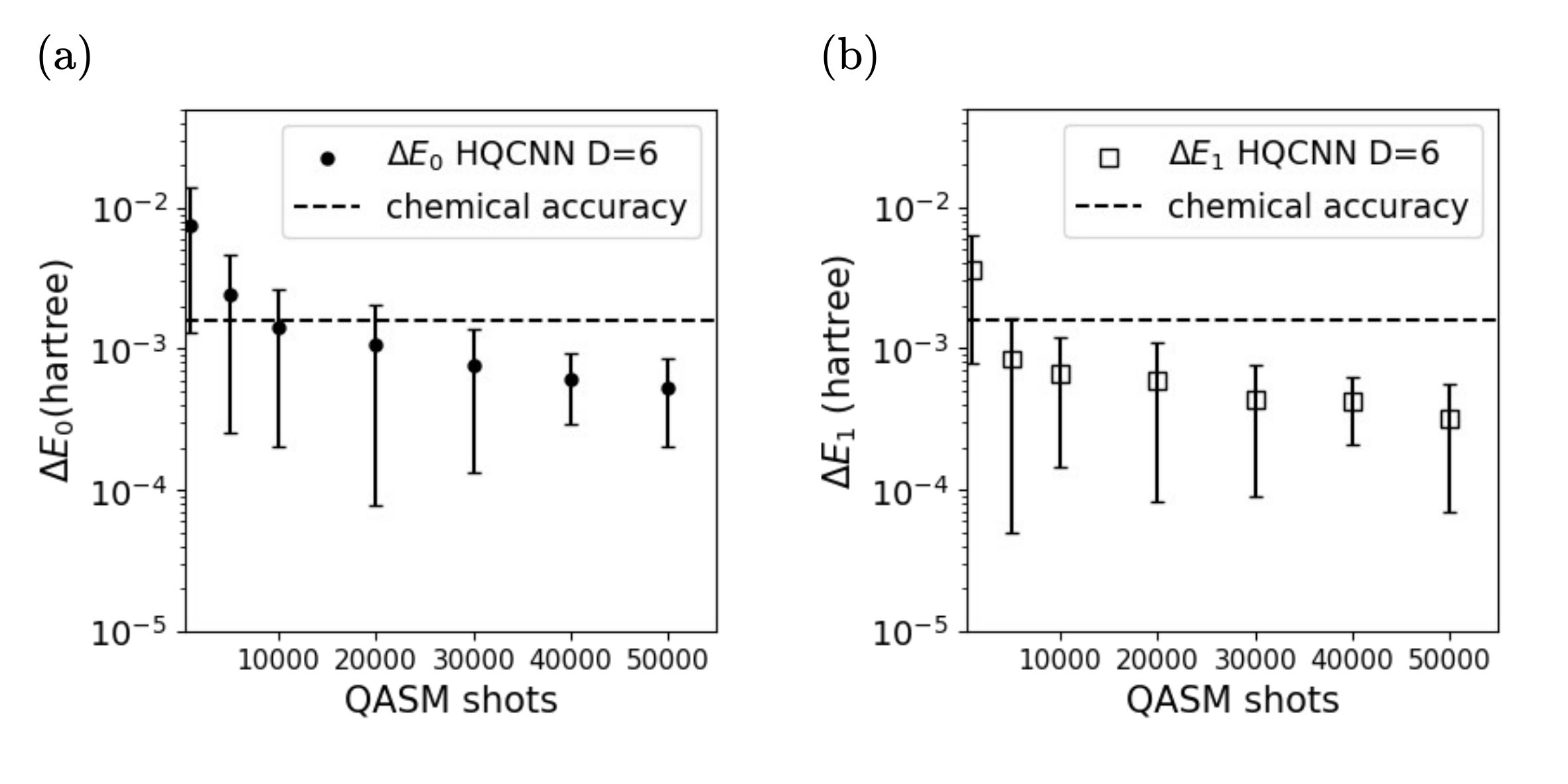}
  \caption{Absolute errors of PES between the HQCNN model 
  and the FCI method at each QASM shot condition. 
  Error bars in the figure represent standard deviation (std) from the mean. 
    The dashed line in the figure is drawn from the value of chemical accuracy, 
    0.001593 hartree. (a) Mean and std of $\Delta E_0$ at each shot condition.
  (b) Mean and std of $\Delta E_1$ at each shot condition.
  } 
  \label{fig6}
\end{figure}

To evaluate quantitatively the sampling effect, a mean value and standard 
deviation (std) of $\Delta E_0$ and $\Delta E_1$ are also calculated.
Figure \ref{fig6} shows the mean and its std of $\Delta E_0$ and $\Delta E_1$ 
at each QASM shot condition. As a result, we find that inference performance 
improves as the QASM shots increase and at least $4\cdot10^4$ shots are required to 
satisfy the chemical accuracy. 

As far as simulations with sampling noise are only concerned, 
our model can predict the energies of both ground and excited 
state with chemical accuracy when a sufficiently large shot number is 
set such as $10^5$. In our HQCNN model, the sampling noise for 
$\langle \sigma_z^i \rangle$ is also considered, therefore, 
the desired precision $\varepsilon$ of the estimation could 
get worse, $\sim \varepsilon(1+ \sum_{i}\frac{\partial E_j}{\partial \langle \sigma_z^i\rangle}  \frac{\langle \sigma_z^i\rangle}{E_j}) $.
According to the previous resource analysis 
\cite{meas}, the total number of necessary measurements $M$ can 
be estimated as $M={K}/{\varepsilon^2}$,
%
%
where 
$K$ is a proportionality constant that depends on the Hamiltonian. 
In the case of \ce{H2} molecule, $K$ is about 0.1. Hence, 
$M >  4\cdot10^{4}$ shots is required for the chemical accuracy 
$\varepsilon\sim 1.6\cdot 10^{-3}$. 
This is roughly consistent with the results in Fig. \ref{fig6}.
%
%
%
%
\section{Discussion}
\label{sec4}
Here we outline open perspectives. In this study, the PQCs that are 
part of the HQCNN model are optimized by using a noiseless statevector 
simulator. In principle, it can also be done on quantum hardware. 
However, the PQCs in this study are never a shallow circuit. As seen 
in the previous VQE study \cite{bp,nbp,lvqe}, a deep circuit based on 
hardware-efficient ansatzes causes troublesome phenomena, known as 
"barren plateaus" where gradients of a cost function vanish exponentially. 
Therefore, to train the HQCNN on quantum hardware efficiently, 
improving the PQC is essential. In addition, development of an efficient 
optimization method for minimizing the cost function \cite{lcost} and an 
error mitigation technique are also important for efficient training 
of the HQCNN model and its highly accurate inference.

As another approach to estimate excited states, variational quantum 
deflation (VQD) \cite{vqd} is well known. From the comparative study \cite{ibe}, 
VQD can exhibit better performance than SSVQE. Hence an HQCNN model 
where VQD is applied instead of SSVQE would enable more accurate prediction of 
energy. However, when VQD is applied to the HQCNN model, 
optimal PQCs for each excited state have to be prepared individually; 
in other words, the optimal parameter set ${\vec{\Theta}}_k$ for 
each excited state $k$ must be determined through a training loop. 
On the other hand, in this SSVQE case, we just have to optimize one 
parameter set $\vec{\Theta}$ even if many excited states exist.

Finally, we refer to an encode layer to transform  classical data to 
quantum data. This study focuses on a simple molecule, \ce{H2}, 
because a simple encode gate, expressed in Eq. (\ref{eq3}), is used. 
However, for a molecular configuration having more atoms, general encode 
layers will be needed. Following standard practices of quantum machine 
learning, ZZ feature map can be applied \cite{qml2}. Further generalized 
encode layers are discussed in a recent study \cite{kiss}.

\section{Conclusions}
\label{sec5}
In summary, we have developed a hybrid quantum-classical neural network to 
predict the PESs of both the ground and the excited state with high accuracy. 
Our model is a surrogate model of the SSVQE in which mutually 
orthogonal initial states are prepared and a cost function for each state is 
calculated by combining PQC and measurements to achieve nonlinear operations.
We have demonstrated that the proposed HQCNN can be trained for PES inference 
of the ground state and the 1st excited state for the \ce{H2} molecule and 
the results can be obtained with chemical accuracy.
Furthermore, we have presented the sampling noise effect on inference 
performance of the proposed model, and in view of that effect, a sufficiently 
large measurement such as $10^5$ is ideally required for chemical accuracy. 
An interesting problem remains, namely, the question of how to improve 
PQC or the classical layer. 
This will be investigated in future work.

%
%

\bibliographystyle{quantum}

%
%

\appendix    
\renewcommand\thefigure{\thesection.\arabic{figure}}
\setcounter{figure}{0}
\setcounter{section}{0} 
\renewcommand{\thesection}{\Alph{section}} 

\section{Optimal parameters of the $(\omega_0,\omega_1)=(1,0)$ HQCNN model}
\label{appenA}
The optimal parameters $(\vec{\theta}^*, \vec{\Theta}^*)$ for the HQCNN model with each 
depth $D$ are summarized as below. 
The parameterized quantum circuits defined in Eq. (\ref{eq4}) and (\ref{eq5}) are illustrated 
in Fig. \ref{fig2}. The optimal parameters 
$\vec{\theta}^*=\{\theta^*_0,\theta^*_1\cdots \theta^*_{nD-1}\}$ 
and $\vec{\Theta}^* = \{\Theta^*_0,\Theta^*_1\cdots \Theta^*_{nD-1}\}$ are fed 
into PQC1 and PQC2 respectively, as shown in Fig. \ref{fig1}.
\subsection{D=2 for 4-qubits PQC}
%
%
The quantum circuit combining PQC1 with Encode1 in Fig. \ref{fig1} is illustrated as follows,
\ \\[0.05cm]
\scalebox{0.8}{
\Qcircuit @C=1.0em @R=0.2em @!R { \\
    \nghost{{q}_{0} :  } & \lstick{{q}_{0} :  } & \gate{\mathrm{H}} & \gate{\mathrm{R_Y}\,(b)} & \ctrl{1} & \gate{\mathrm{R_Y}\,(\theta^*_0)} & \qw \barrier[0em]{3} & \qw & \ctrl{1} & \gate{\mathrm{R_Y}\,(\theta^*_4)} & \qw \barrier[0em]{3} & \qw & \qw & \qw\\
    \nghost{{q}_{1} :  } & \lstick{{q}_{1} :  } & \gate{\mathrm{H}} & \gate{\mathrm{R_Y}\,(b)} & \targ & \ctrl{1} & \gate{\mathrm{R_Y}\,(\theta^*_1)} & \qw & \targ & \ctrl{1} & \gate{\mathrm{R_Y}\,(\theta^*_5)} & \qw & \qw & \qw\\
    \nghost{{q}_{2} :  } & \lstick{{q}_{2} :  } & \gate{\mathrm{H}} & \gate{\mathrm{R_Y}\,(b)} & \ctrl{1} & \targ & \gate{\mathrm{R_Y}\,(\theta^*_2)} & \qw & \ctrl{1} & \targ & \gate{\mathrm{R_Y}\,(\theta^*_6)} & \qw & \qw & \qw\\
    \nghost{{q}_{3} :  } & \lstick{{q}_{3} :  } & \gate{\mathrm{H}} & \gate{\mathrm{R_Y}\,(b)} & \targ & \gate{\mathrm{R_Y}\,(\theta^*_3)} & \qw & \qw & \targ & \gate{\mathrm{R_Y}\,(\theta^*_7)} & \qw & \qw & \qw & \qw\\
\\ }}
\ \\[0.05cm]
The quantum circuit combining PQC2 with Encode2 in Fig. \ref{fig1} is illustrated as follows,
\ \\[0.05cm]
\scalebox{0.8}{
\Qcircuit @C=1.0em @R=0.2em @!R { \\
    \nghost{{q}_{0} :  } & \lstick{{q}_{0} :  } & \gate{\mathrm{H}} & \gate{\mathrm{R_Y}\,(b_0')} & \ctrl{1} & \gate{\mathrm{R_Y}\,(\Theta^*_0)} & \qw \barrier[0em]{3} & \qw & \ctrl{1} & \gate{\mathrm{R_Y}\,(\Theta^*_4)} & \qw \barrier[0em]{3} & \qw & \qw & \qw\\
    \nghost{{q}_{1} :  } & \lstick{{q}_{1} :  } & \gate{\mathrm{H}} & \gate{\mathrm{R_Y}\,(b_1')} & \targ & \ctrl{1} & \gate{\mathrm{R_Y}\,(\Theta^*_1)} & \qw & \targ & \ctrl{1} & \gate{\mathrm{R_Y}\,(\Theta^*_5)} & \qw & \qw & \qw\\
    \nghost{{q}_{2} :  } & \lstick{{q}_{2} :  } & \gate{\mathrm{H}} & \gate{\mathrm{R_Y}\,(b_2')} & \ctrl{1} & \targ & \gate{\mathrm{R_Y}\,(\Theta^*_2)} & \qw & \ctrl{1} & \targ & \gate{\mathrm{R_Y}\,(\Theta^*_6)} & \qw & \qw & \qw\\
    \nghost{{q}_{3} :  } & \lstick{{q}_{3} :  } & \gate{\mathrm{H}} & \gate{\mathrm{R_Y}\,(b_3')} & \targ & \gate{\mathrm{R_Y}\,(\Theta^*_3)} & \qw & \qw & \targ & \gate{\mathrm{R_Y}\,(\Theta^*_7)} & \qw & \qw & \qw & \qw\\
\\ }}
\ \\[0.1cm]
where, $\{b_i'\} =\{\pi\langle\sigma^i_z\rangle\}$. 
Then, the parameters $(\vec{\theta}^*, \vec{\Theta}^*)$ is summarized in the following table.
The parameter-index $i$ is incremented according to the order of the block 
circuit in Fig \ref{fig2}, i.e., the angle of $R_y$ rotation gate on the $i$th 
qubit of $D$th block circuit is given by $\theta^*_{i+4(D-1)}$ for PQC1 
($\Theta^*_{i+4(D-1)}$ for PQC2).
\ \\[0.5cm]
 \tablefirsthead{%
 \toprule
 $i$ & $\theta^*_i$ & $\Theta^*_i$  
 \tabularnewline
 \midrule}
 \tabletail{\midrule}
\tablelasttail{\bottomrule}
\begin{supertabular}{ccc}
  \begin{filecontents*}{d_2.csv}
Nr;x;y
0; 0.44990398;  -0.07895043
1; 0.00492587;  -2.43631272
2; 1.42098544;  2.46370587
3; 0.60825627;  0.26725636
4; -0.47728727; -0.00459582
5; 1.60923202;  -1.55279297
6; 0.38932374;  -0.24098644
7; 0.3966547; 1.51973278
\end{filecontents*}
\csvreader[ separator=semicolon,
 late after line=\\,
 ]{d_2.csv}{1=\Nr,2=\x,3=\y}{\Nr&\x&\y}
 \bottomrule
 \end{supertabular}
\subsection{D=3 for 4-qubits PQC}
 \tablefirsthead{%
 \toprule
 $i$ & $\theta^*_i$ & $\Theta^*_i$  
 \tabularnewline
 \midrule}
 \tabletail{\midrule}
\tablelasttail{\bottomrule}
\begin{supertabular}{ccc}
  \begin{filecontents*}{d_3.csv}
Nr;x;y
0  ;   1.96276914e+00 ;  -1.12944101e-03
1  ;   1.25670686e+00 ;  -5.46716619e-01
2  ;   3.29694642e-01 ;   8.29253916e-01
3  ;   1.30778739e+00 ;   1.16291385e-01
4  ;  -8.11410342e-01 ;  -2.53018795e-03
5  ;  -2.13754993e-01 ;   6.23986909e-03
6  ;   2.40197722e-01 ;   4.50162758e-01
7  ;   1.41471586e+00 ;   6.19085465e-02
8  ;   7.42486302e-01 ;  -9.55612005e-04
9  ;   1.53853476e+00 ;   5.00809978e-01
10 ;  -1.01820297e-01 ;  -8.19532007e-01
11 ;   1.62537711e-01 ;   1.58511488e+00
\end{filecontents*}
\csvreader[ separator=semicolon,
 late after line=\\,
 ]{d_3.csv}{1=\Nr,2=\x,3=\y}{\Nr&\x&\y}
 \bottomrule
 \end{supertabular}
\subsection{D=4 for 4-qubits PQC}
 \tablefirsthead{%
 \toprule
 $i$ & $\theta^*_i$ & $\Theta^*_i$  
 \tabularnewline
 \midrule}
 \tabletail{\midrule}
\tablelasttail{\bottomrule}
\begin{supertabular}{ccc}
  \begin{filecontents*}{d_4.csv}
Nr;x;y
0 ;    1.18244917     ;    1.56155777
1 ;    -0.03800419    ;    -0.02127339
2 ;    0.33551713     ;      1.55263955
3 ;     0.63058047    ;    1.70198366
4 ;     2.13454417    ;    1.56133918
5 ;    1.44831056     ;     -0.00365438
6 ;      1.03974931   ;     0.00587906
7 ;     0.08796629    ;    0.02509409
8 ;     2.08712489    ;     -0.24766553
9 ;     1.12208658    ;    1.31541851
10;     -0.52375853   ;    2.20279135
11;      0.77828267   ;    -0.06444451
12;      -0.97212089  ;    -1.55945777
13;     0.13481862    ;    0.00841186
14;     0.81579652    ;    -0.00459007
15;      0.42064851   ;    1.51055261
\end{filecontents*}
\csvreader[ separator=semicolon,
 late after line=\\,
 ]{d_4.csv}{1=\Nr,2=\x,3=\y}{\Nr&\x&\y}
 \bottomrule
 \end{supertabular}
\begin{figure}[h]
  \centering
  \includegraphics[scale=0.57]{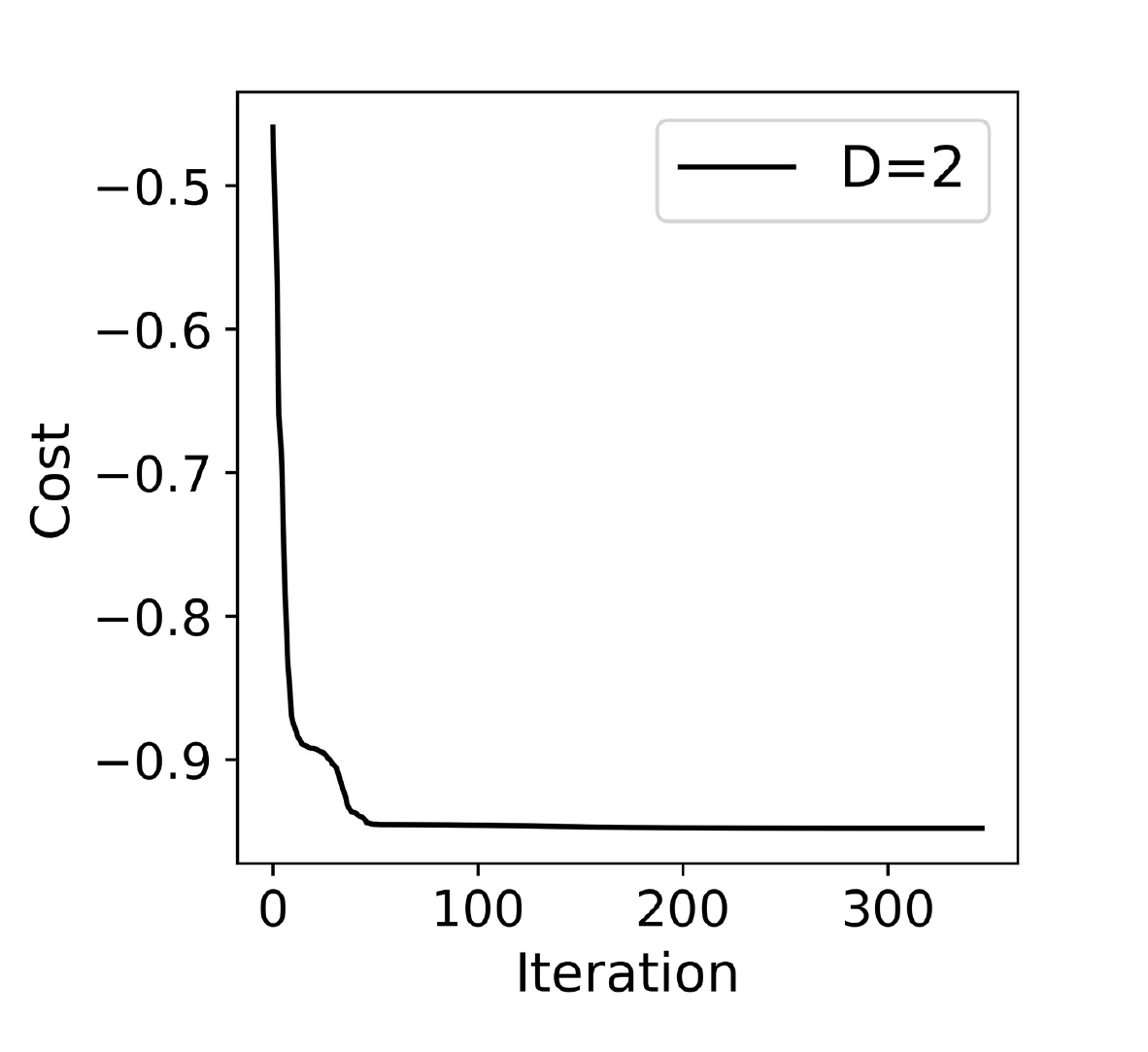}
  \includegraphics[scale=0.57]{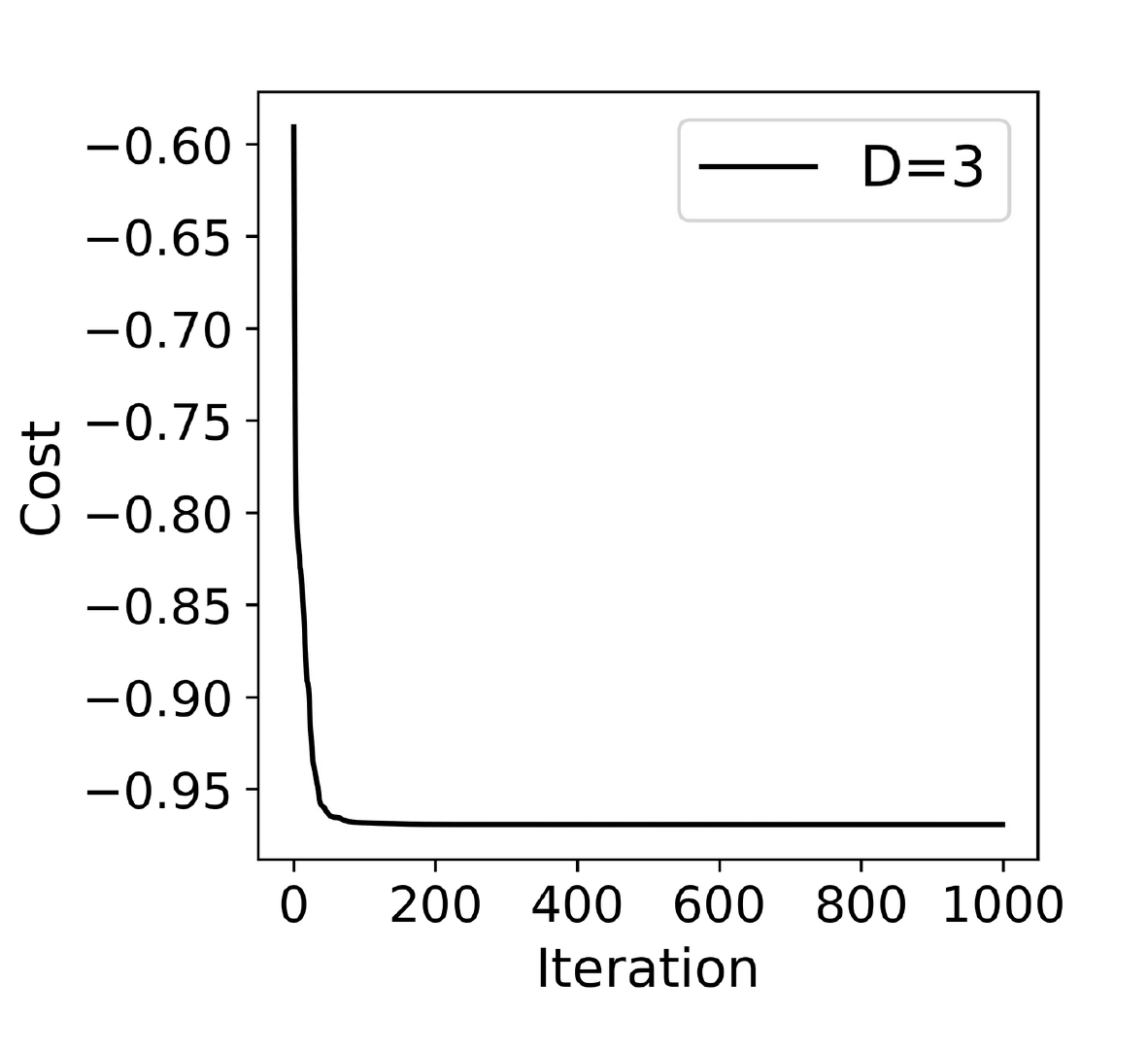}
  \includegraphics[scale=0.57]{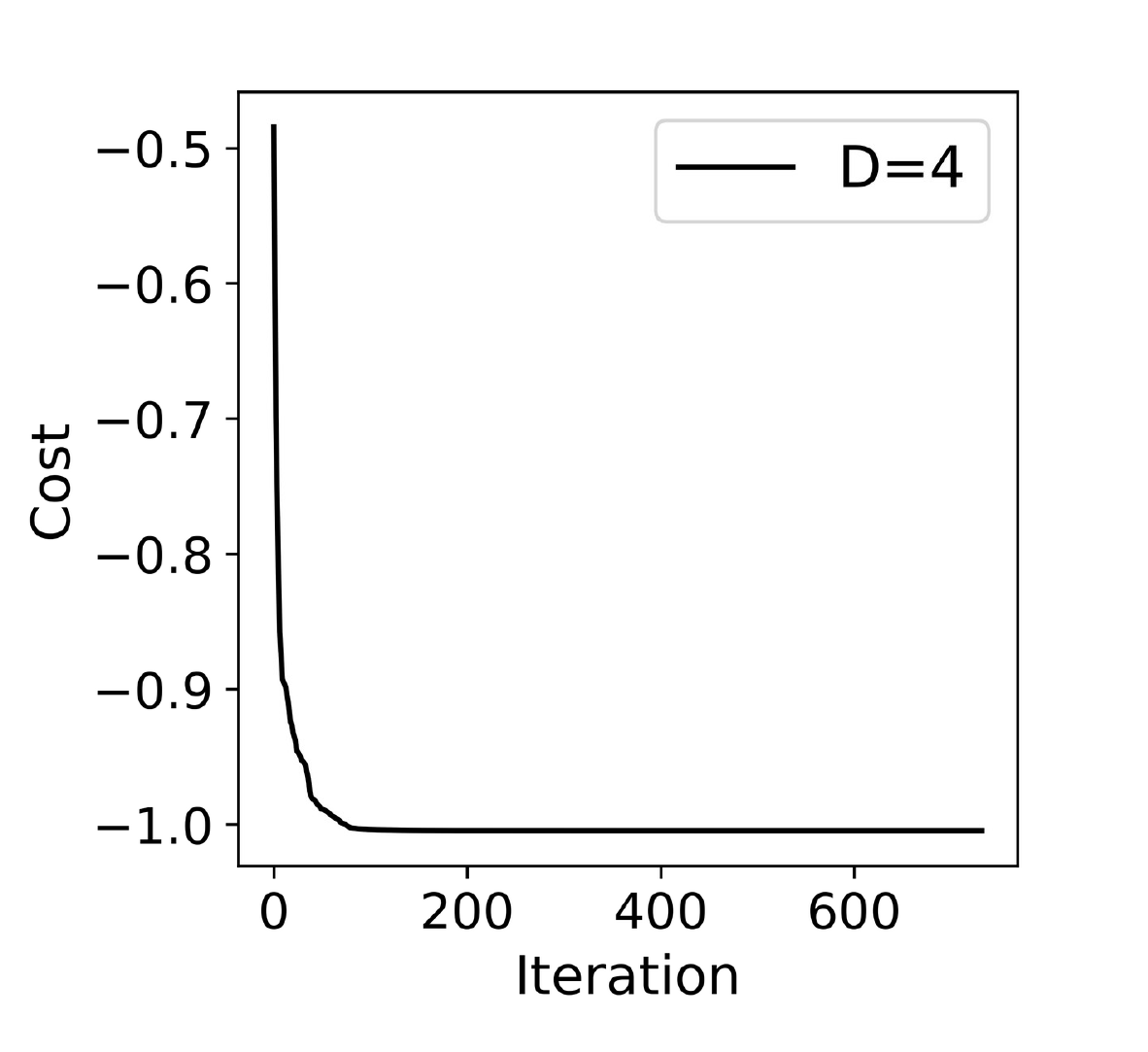}
  \includegraphics[scale=0.57]{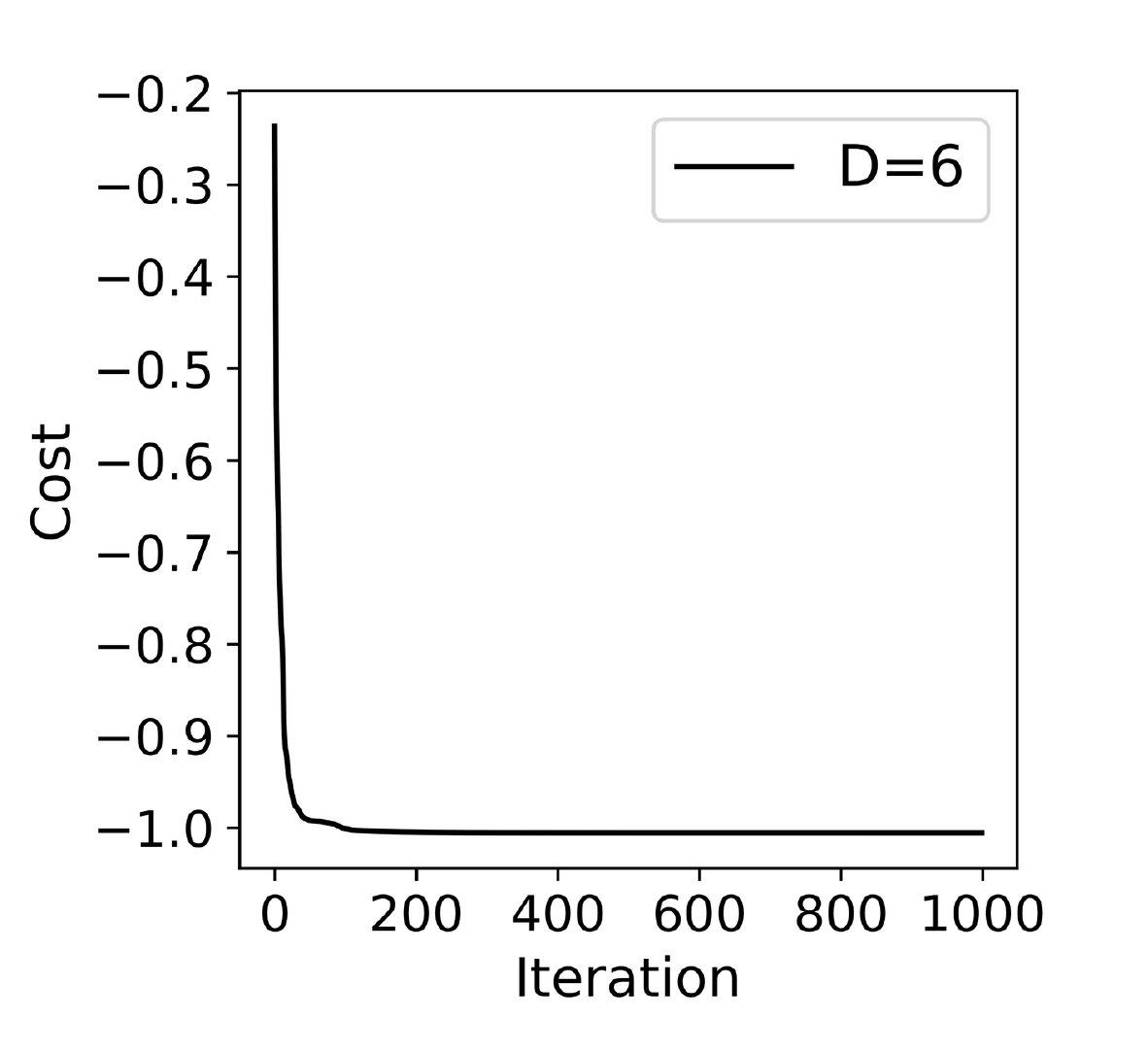}
  \caption{Optimization process for each $D$ case using BFGS optimizer. The computational 
  conditions are summarized in Sec. \ref{sec3-5}} 
\end{figure}
\subsection{D=6 for 4-qubits PQC}
 \tablefirsthead{%
 \toprule
 $i$ & $\theta^*_i$ & $\Theta^*_i$  
 \tabularnewline
 \midrule}
 \tabletail{\midrule}
\tablelasttail{\bottomrule}
\begin{supertabular}{ccc}
  \begin{filecontents*}{d_6.csv}
Nr;x;y
0  ;  1.96944586e-01    ;    2.16488379e-01
1  ;  6.96085135e-01    ;   -1.47369675e+00
2  ;  2.75194632e-01    ;   -6.09809058e-01
3  ;  1.13995330e+00    ;   2.18701015e+00
4  ;   7.98447448e-01   ;    4.54891917e-02
5  ;  9.27048824e-02    ;   -6.58927429e-02
6  ;  1.39434016e+00    ;   -3.64902005e-02
7  ;  2.45081388e-01    ;   -6.52573592e-01
8  ;   5.50966511e-02   ;     5.19154392e-04
9  ;  -1.20161139e+00   ;   3.60913707e-01
10 ;   5.01887217e-01   ;   -1.80183282e-01
11 ;   4.12684265e-01   ;   1.54652908e+00
12 ;   1.05144072e+00   ;     5.44473649e-04
13 ;   1.87287163e-01   ;   4.70659767e-01
14 ;   -2.61354669e-01  ;   -1.73169717e-01
15 ;   6.14793737e-01   ;   -1.11815028e+00
16 ;   2.46727571e-01   ;     4.58367070e-04
17 ;   -4.15112509e-02  ;   3.04351803e-01
18 ;   3.09647969e-01   ;   3.17354653e-01
19 ;   1.68123352e+00   ;   -3.38641312e-01
20 ;    1.49119866e-01  ;    -2.13794185e-03
21 ;   1.90244017e-01   ;   2.44324411e-01
22 ;   1.47124686e+00   ;   -2.27303510e+00
23 ;   1.19942530e+00   ;    2.22905860e-02
\end{filecontents*}
\csvreader[ separator=semicolon,
 late after line=\\,
 ]{d_6.csv}{1=\Nr,2=\x,3=\y}{\Nr&\x&\y}
 \bottomrule
 \end{supertabular}
%
\section{Results with and without classical layer}
\renewcommand\thefigure{\thesection.\arabic{figure}}
\setcounter{figure}{0}
\label{appenB}
The results of PES estimation with and without classical layer (C-layer) 
are displayed in Fig \ref{fig8}. The parameterized quantum circuit without C-layer 
corresponds to the PQC1 with $D=8$, in which the number of parameters $\{\vec{\theta}^* \}$ is 
32.
\ \\[0.1cm]
\scalebox{0.77}{
\Qcircuit @C=1.0em @R=0.2em @!R { \\
    \nghost{{q}_{0} :  } & \lstick{{q}_{0} :  } & \gate{\mathrm{H}} & \gate{\mathrm{R_Y}\,(b)} & \ctrl{1} & \gate{\mathrm{R_Y}\,(\theta^*_0)} & \qw \barrier[0em]{3} & \qw & \ctrl{1} & \gate{\mathrm{R_Y}\,(\theta^*_4)} & \qw \barrier[0em]{3} & \qw & \ctrl{1} & \gate{\mathrm{R_Y}\,(\theta^*_8)} & \qw \barrier[0em]{3} & \qw & \qw\\ 
    \nghost{{q}_{1} :  } & \lstick{{q}_{1} :  } & \gate{\mathrm{H}} & \gate{\mathrm{R_Y}\,(b)} & \targ & \ctrl{1} & \gate{\mathrm{R_Y}\,(\theta^*_{1})} & \qw & \targ & \ctrl{1} & \gate{\mathrm{R_Y}\,(\theta^*_{5})} & \qw & \targ & \ctrl{1} & \gate{\mathrm{R_Y}\,(\theta^*_{9})} & \qw & \qw\\
    \nghost{{q}_{2} :  } & \lstick{{q}_{2} :  } & \gate{\mathrm{H}} & \gate{\mathrm{R_Y}\,(b)} & \ctrl{1} & \targ & \gate{\mathrm{R_Y}\,(\theta^*_{2})} & \qw & \ctrl{1} & \targ & \gate{\mathrm{R_Y}\,(\theta^*_{6})} & \qw & \ctrl{1} & \targ & \gate{\mathrm{R_Y}\,(\theta^*_{10})} & \qw & \qw\\
    \nghost{{q}_{3} :  } & \lstick{{q}_{3} :  } & \gate{\mathrm{H}} & \gate{\mathrm{R_Y}\,(b)} & \targ & \gate{\mathrm{R_Y}\,(\theta^*_{3})} & \qw & \qw & \targ & \gate{\mathrm{R_Y}\,(\theta^*_{7})} & \qw & \qw & \targ & \gate{\mathrm{R_Y}\,(\theta^*_{11})} & \qw & \qw & \qw 
    \\ }
}
\ \\[0.3cm]
\scalebox{0.77}{
\Qcircuit @C=1.0em @R=0.2em @!R { \\
    \nghost{{q}_{0} :  } & \ctrl{1} & \gate{\mathrm{R_Y}\,(\theta^*_{12})} & \qw \barrier[0em]{3} & \qw & \ctrl{1} & \gate{\mathrm{R_Y}\,(\theta^*_{16})} & \qw \barrier[0em]{3} & \qw & \ctrl{1} & \gate{\mathrm{R_Y}\,(\theta^*_{20})} & \qw \barrier[0em]{3} & \qw & \qw\\ 
    \nghost{{q}_{1} :  } & \targ & \ctrl{1} & \gate{\mathrm{R_Y}\,(\theta^*_{13})} & \qw & \targ & \ctrl{1} & \gate{\mathrm{R_Y}\,(\theta^*_{17})} & \qw & \targ & \ctrl{1} & \gate{\mathrm{R_Y}\,(\theta^*_{21})} & \qw & \qw\\
    \nghost{{q}_{2} :  } & \ctrl{1} & \targ & \gate{\mathrm{R_Y}\,(\theta^*_{14})} & \qw & \ctrl{1} & \targ & \gate{\mathrm{R_Y}\,(\theta^*_{18})} & \qw & \ctrl{1} & \targ & \gate{\mathrm{R_Y}\,(\theta^*_{22})} & \qw & \qw\\
    \nghost{{q}_{3} :  } & \targ & \gate{\mathrm{R_Y}\,(\theta^*_{15})} & \qw & \qw & \targ & \gate{\mathrm{R_Y}\,(\theta^*_{19})} & \qw & \qw & \targ & \gate{\mathrm{R_Y}\,(\theta^*_{23})} & \qw & \qw & \qw 
\\ }
}
\ \\[0.3cm]
\scalebox{0.77}{
\Qcircuit @C=1.0em @R=0.2em @!R { \\
    \nghost{{q}_{0} :  } & \ctrl{1} & \gate{\mathrm{R_Y}\,(\theta^*_{24})} & \qw \barrier[0em]{3} & \qw & \ctrl{1} & \gate{\mathrm{R_Y}\,(\theta^*_{28})} & \qw \barrier[0em]{3} & \qw & \qw \\ 
    \nghost{{q}_{1} :  } & \targ & \ctrl{1} & \gate{\mathrm{R_Y}\,(\theta^*_{25})} & \qw & \targ & \ctrl{1} & \gate{\mathrm{R_Y}\,(\theta^*_{29})} & \qw & \qw\\
    \nghost{{q}_{2} :  } & \ctrl{1} & \targ & \gate{\mathrm{R_Y}\,(\theta^*_{26})} & \qw & \ctrl{1} & \targ & \gate{\mathrm{R_Y}\,(\theta^*_{30})} & \qw & \qw\\
    \nghost{{q}_{3} :  } & \targ & \gate{\mathrm{R_Y}\,(\theta^*_{27})} & \qw & \qw & \targ & \gate{\mathrm{R_Y}\,(\theta^*_{31})} & \qw & \qw & \qw \\
\\ }
}

\ \\[0.5cm]
The optimal parameters $\theta^*_i$ are summarized as below.
\ \\[0.1cm]

 \tablefirsthead{%
 \toprule
 $i$ & $\theta^*_i$   
 \tabularnewline
 \midrule}
 \tabletail{\midrule}
\tablelasttail{\bottomrule}
\begin{supertabular}{ccc}
  \begin{filecontents*}{d_4_no_meas.csv}
Nr;x
0 ;  -2.21923412
1 ;  1.75397721
2 ;  -0.11685527
3 ;  1.04819334
4 ;  -0.60779518
5 ;  -0.03740226
6 ;    0.88396476
7 ;   0.44289291
8 ;  -0.04146204
9 ;   1.19623164
10;   0.49826975
11;   0.06755827
12;    -0.07466358
13;   0.61787473
14;   0.95700257
15;    -0.94629637
16;   1.46049613
17;   -0.3841755
18;     0.77517227
19;    1.31200736
20;    1.56461399
21;    1.43727198
22;   1.61064495
23;   2.16997706
24;    -0.05908558
25;   1.14194436
26;   -0.00621807
27;    0.79989797
28;   -0.58098377
29;    0.09029734
30;    0.0309237
31;     1.65240314
\end{filecontents*}
\csvreader[ separator=semicolon,
 late after line=\\,
 ]{d_4_no_meas.csv}{1=\Nr,2=\x}{\Nr&\x}
 \bottomrule
 \end{supertabular}
\ \\[0.3cm]
\begin{figure}[h]
  \centering
  \includegraphics[scale=0.57]{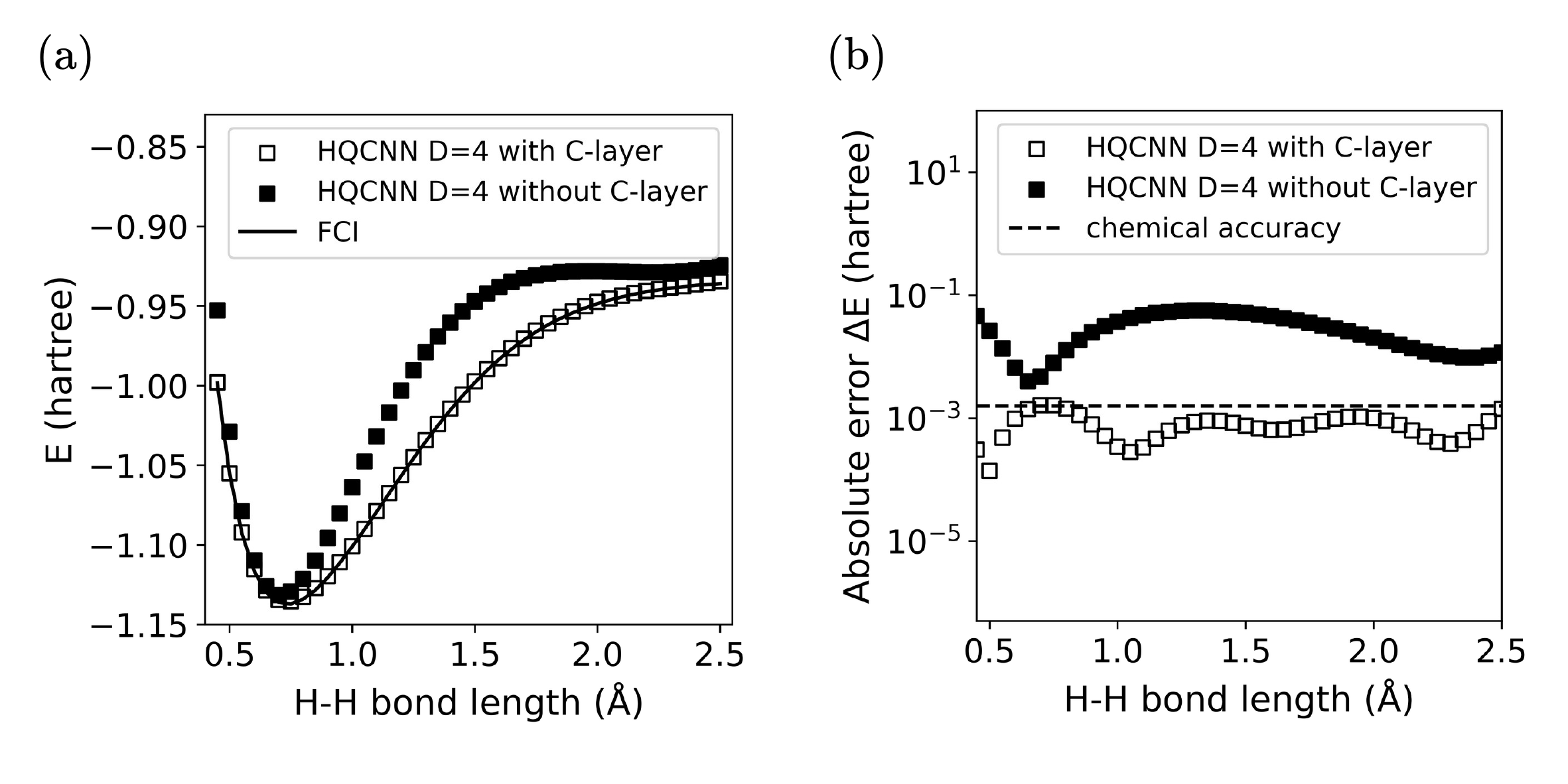}
  \caption{(a) PES estimation for the \ce{H2} molecule using the $D=4$ HQCNN model 
  with and without classical layer (C-layer) discussed in Sec. \ref{sec3-3}.
  (b) Absolute error of the energies between each HQCNN and the FCI method. 
  The dotted line in the figure is drawn from the value of chemical 
  accuracy, 0.001593 hartree.
  } 
  \label{fig8}
\end{figure}
\newpage
\section{Optimal parameters of the $(\omega_0,\omega_1)=(1,0.5)$ HQCNN model}
\renewcommand\thefigure{\thesection.\arabic{figure}}
\setcounter{figure}{0}
\label{appenC}
\subsection{D=6 for 4-qubits PQC}
The optimal parameters $(\vec{\theta}^*, \vec{\Theta}^*)$ in Sec. \ref{sec3-2} 
are summarized in the following table.
\ \\[0.5cm]
 \tablefirsthead{%
 \toprule
 $i$ & $\theta^*_i$ & $\Theta^*_i$  
 \tabularnewline
 \midrule}
 \tabletail{\midrule}
\tablelasttail{\bottomrule}
\begin{supertabular}{ccc}
  \begin{filecontents*}{d_6_2.csv}
Nr;x;y
0  ;  4.06555514e-01     ;    3.23397197e-01
1  ;  3.90765681e-01     ;   3.14126164e+00
2  ;   1.28552322e-01    ;  -5.20690613e-05
3  ;  -2.03348271e-01    ;   4.99569944e-01
4  ;   8.52936329e-01    ;    1.57060921e+00
5  ;   1.09099632e+00    ;   7.56231356e-05
6  ;  -2.15279649e-02    ;  -3.63973378e-04
7  ;  3.19553075e-01     ;   1.57068465e+00
8  ;   5.42824781e-01    ;   -8.81188684e-04
9  ;  1.97727918e-01     ;   1.44990590e+00
10 ;   3.93177448e-01    ;   9.48805894e-01
11 ;   2.89423839e-01    ;   -2.16437280e-04
12 ;    4.57329007e-01   ;    1.50413458e-04
13 ;  -1.11312219e+00    ;  1.57060129e+00
14 ;   1.60299414e+00    ;   -4.96648893e-01
15 ;    9.58759382e-01   ;   1.57073176e+00
16 ;   -6.38281689e-01   ;    1.57087092e+00
17 ;   1.43562988e-01    ;  1.57076878e+00
18 ;    1.39205790e+00   ;  6.21569280e-01
19 ;  1.82419115e+00     ;  -1.62844482e+00
20 ;    4.95991802e-02   ;    1.54094136e-05
21 ;    6.24543518e-02   ;    1.33004098e-05
22 ;   -3.43793502e-01   ;   1.57074102e+00
23 ;    7.32904855e-01   ;  1.17479292e-04
\end{filecontents*}
\csvreader[ separator=semicolon,
 late after line=\\,
 ]{d_6_2.csv}{1=\Nr,2=\x,3=\y}{\Nr&\x&\y}
 \bottomrule
 \end{supertabular}
\ \\[0.3cm]
\begin{figure}[h]
  \includegraphics[scale=0.57]{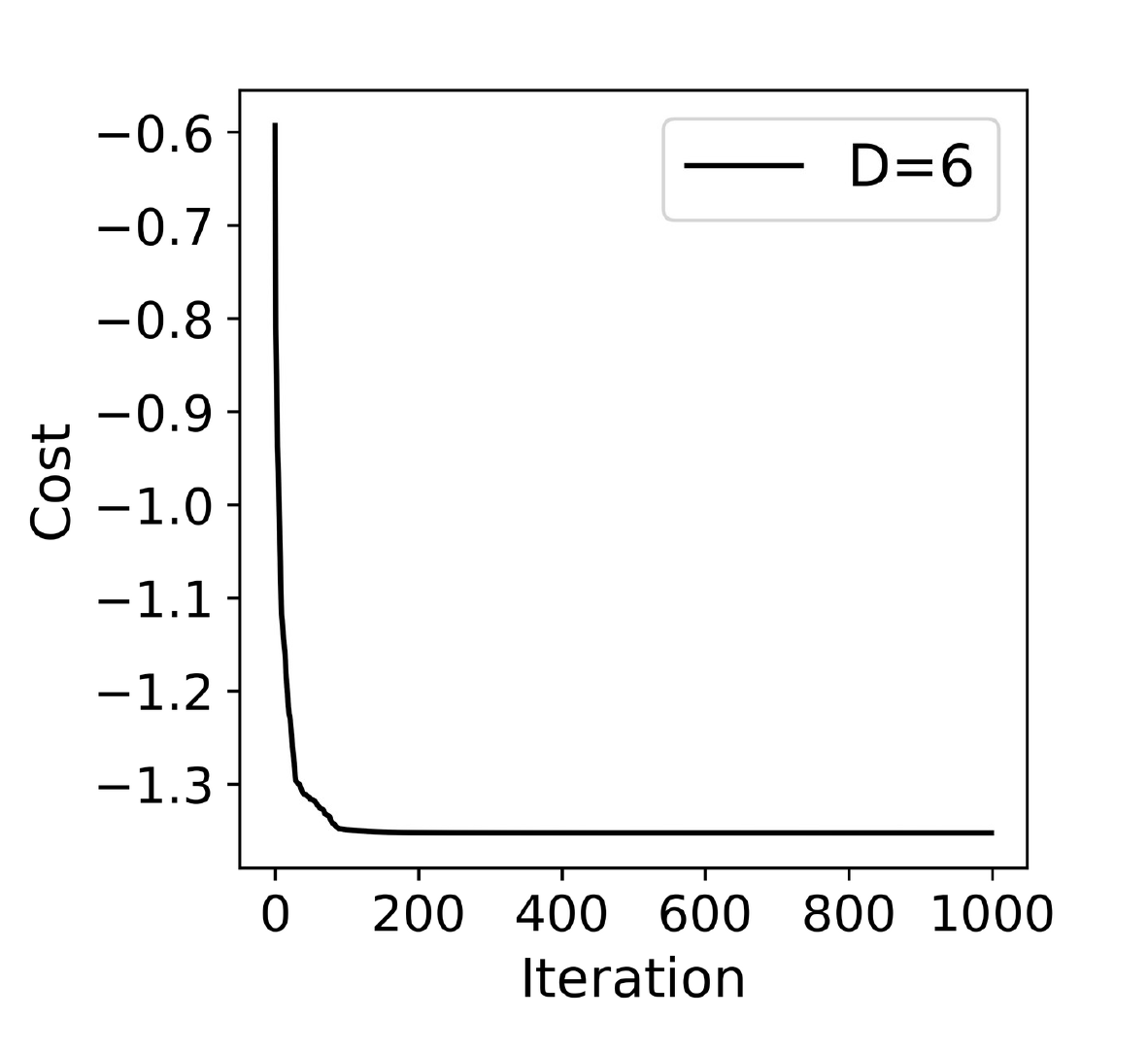}
  \caption{Optimization process for $D=6$ case using BFGS optimizer.
  The computational conditions are summarized in Sec. \ref{sec3-5}} 
\end{figure}
\end{document}